\documentclass[10pt, conference]{IEEEtran}
\usepackage{listings}
\usepackage{graphicx,color}
\usepackage{booktabs}
\usepackage{longtable}
\usepackage{subfigure}
\usepackage{amsmath,amssymb,amsfonts}
\usepackage{algorithmic}
\usepackage{graphicx}
\usepackage{textcomp}
\usepackage{xcolor}
\usepackage{multirow}
\usepackage{cite}
\usepackage{multirow}
\usepackage{url}

\def\etal{{\it et al.}\hspace{0.1pc}}

\def\BibTeX{{\rm B\kern-.05em{\sc i\kern-.025em b}\kern-.08em
    T\kern-.1667em\lower.7ex\hbox{E}\kern-.125emX}}

\makeatletter
\def\footnoterule{\relax%
  \kern-5pt
  \hbox to \columnwidth{\hfill\vrule width \columnwidth height 0.4pt\hfill}
  \kern4.6pt}
\makeatother
  
\IEEEoverridecommandlockouts

\begin{document}

\title{Demystifying the Nvidia Ampere Architecture through Microbenchmarking and Instruction-level Analysis}

\author{
\IEEEauthorblockN{Hamdy Abdelkhalik}
\IEEEauthorblockA{
\textit{New Mexico State University}\\
enghamdy@nmsu.edu \vspace{-20pt}}
\and
\IEEEauthorblockN{Yehia Arafa * }
\IEEEauthorblockA{
% \textit{Qualcomm Technologies, Inc.}\\
% yarafa@qti.qualcomm.com \vspace{-20pt}}
% \textit{Qualcomm Technologies, Inc.,}\\
\textit{New Mexico State University}\\
yarafa@nmsu.edu \vspace{-20pt}}

\and
\IEEEauthorblockN{Nandakishore Santhi}
\IEEEauthorblockA{
\textit{Los Alamos National Laboratory}\\
nsanthi@lanl.gov \vspace{-20pt}}\\
\and
\IEEEauthorblockN{Abdel-Hameed Badawy }
\IEEEauthorblockA{
\textit{New Mexico State University}\\
badawy@nmsu.edu \vspace{-20pt}}

% \thanks{\hline}
\thanks{* Now with Qualcomm Technologies, Inc.}
}

\maketitle

\begin{abstract}

Graphics processing units (GPUs) are now considered the leading hardware to accelerate general-purpose workloads such as AI, data analytics, and HPC. Over the last decade, researchers have focused on demystifying and evaluating the microarchitecture features of various GPU architectures beyond what vendors reveal. This line of work is necessary to understand the hardware better and build more efficient workloads and applications. Many works have studied the recent Nvidia architectures, such as Volta and Turing, comparing them to their successor, Ampere. However, some microarchitecture features, such as the clock cycles for the different instructions, have not been extensively studied for the Ampere architecture. 
In this paper, we study the clock cycles per instructions with various data types found in the instruction-set architecture (ISA) of Nvidia GPUs. Using microbenchmarks, we measure the clock cycles for PTX ISA instructions and their SASS ISA instructions counterpart. we further calculate the clock cycle needed to access each memory unit. We also demystify the new version of the tensor core unit found in the Ampere architecture by using the WMMA API and measuring its clock cycles per instruction and throughput for the different data types and input shapes. 
The results found in this work should guide software developers and hardware architects. Furthermore, the clock cycles per instructions are widely used by performance modeling simulators and tools to model and predict the performance of the hardware.

\end{abstract}


\section{Introduction}

Graphics processing units (GPUs) have significantly increased in accelerating general-purpose applications from neural networks to scientific computing. GPUs are now considered the main hardware component in any high-performance supercomputer. For instance, Meta built one of the fastest supercomputers based on Nvidia Ampere architecture GPUs (A100)~\cite{fb_supercomputer}, and they are extending it to be the most powerful supercomputer in the world by mid-2022. Besides, tens of the top500 supercomputers~\cite{top500} are GPU-accelerated.

Nvidia provides a new architecture generation with updated features every two years with little micro-architecture information about these features, making it difficult to quantify. This raises the need to study the effect of new features on the performance of applications. Nvidia introduced the tensor core (TC) unit to accelerate deep neural networks with the introduction of Volta. This version of TC operates on FP16 and FP32 precision operations. Ampere architectures added a sparsity feature and new data precisions for the TC such as Int8, Int4, FP64, bf16, and TF32. Usually, there is little information beyond what the vendors choose to reveal in their whitepapers, which raises the need to quantify these features. Thus, researchers try to explain and demystify the new features of each GPU generation~\cite{zhang2017understanding,ho2017exploiting,lustig2019formal,wong2010demystifying,jia2018dissecting,jia2019dissecting}. However, some areas still have not been fully covered in the literature. In this work, we focus on demystifying the clock cycle latency at the granularity of the instructions in the Instruction-set architecture (ISA). Similar work has been proposed before. For instance, the authors in~\cite{van2022isolating} adapted some microbenchmarks to demystify some hardware units such as the memory, the tensor cores, and the arithmetic units. However, they only calculated the latency for memories at the granularity of the warp and the block, not the instructions. In another work~\cite{andersch2015latency}, the authors calculated the latencies for only the memory hierarchy of older generations. 

This paper presents microbenchmark analyses to dissect the instruction clock cycles per instructions for the Nvidia Ampere GPU architecture~\cite{Nvidia_ampere_arch_doc}. The microbenchmarks presented in this work are based on \textit{Parallel Thread Execution} (PTX)~\cite{Nvidia_PTX}. PTX is an intermediate representation between the high-level language (CUDA) and the assembly language (SASS). So, it is portable among different architectures. PTX is open-source and well documented. However, its instructions do not directly execute on the hardware. It has to be converted to another architecture-dependent ISA. SASS, in this case, is closed-source and compatible only within each architecture family. This paper shows how each PTX instruction is mapped to SASS instruction while measuring the clock cycles for both ISAs. Furthermore, we present the clock cycle needed to access each memory unit. The microbenchmarks are based on a previous work by Arafa~\etal~\cite{arafa2019low}, which calculated the clock cycle latencies for various instructions on different Nvidia architectures. However, there are no such studies done on the Ampere architecture. We also show the clock cycles and throughput for tensor core instructions on different data types. 

Measuring the instructions clock cycles helps predict performance by GPU modeling tools. For instance, Arafa~\etal~~\cite{arafa2019ppt} showed that by adopting correct latency for the GPU instructions, their performance model can improve its prediction accuracy compared to the actual hardware. Furthermore, Andersch~\etal~\cite{andersch2015latency} have proven the critical relationship between the latencies and the performance. This work is the first step in accurately modeling the Ampere GPU architecture.

The main \textbf{contributions} of this paper are as follows:
\begin{itemize}

\item  We demystify the Nvidia Ampere~\cite{Nvidia_ampere_arch_doc} GPU architecture through microbenchmarking by measuring the clock cycles latency per instruction on different data types.
\item We show the mapping of PTX instructions to the sass instructions while measuring the clock cycles for both.
\item We calculate the Ampere architecture tensor cores instructions (\textit{WMMA}) clock cycle latency and throughput while clarifying their PTX and SASS instructions. 
\item We measure the access latency of the different memory units inside the Ampere architecture. 
\end{itemize}

%====================================================
%====================================================
\section{Background}
\label{sec:background}

Unlike multi-core CPUs, which have several powerful processors, GPUs have tens of simple processors that can work simultaneously to perform a specific task efficiently. This is viable for many applications that require numerous work to be performed in parallel, such as artificial intelligence and scientific computing.

An Nvidia GPU consists of several streaming multiprocessors (SM). The number of SMs varies with the generation of the GPU. Older architectures, such as Kepler, have fewer SMs (15 or 24), while contemporary architectures, such as Ampere, have a more significant number (124). The computation resources inside the SMs also vary depending on the architecture generation. Each SM is divided into hundreds of small cores performing different operations. GPUs have different types of memory units. The global memory and L2 cache are shared with all SMs. Furthermore, it has L1 caches, which are private to each SM. Moreover, threads inside a block can communicate through the shared memory.

The need for GPUs in many essential fields nowadays forces the vendors to enhance their GPU architecture to provide better performance. NVIDIA provides a new architecture every two years. New architectures not only have new hardware units but also may contain new ISA that increases performance. For example, in the Ampere architecture, Nvidia introduced much enhancement in the tensor core unit, making it faster and run on larger matrices. Moreover, it introduced the new L2 cache residency control feature, which automatically manages data to keep or evict from the cache. 

Although these features are well documented in the whitepaper and online review websites, there are little information on the microarchitecture and the instruction-level enhancements found in the recent Ampere architecture. This paper fills this gap by providing a detailed instruction-level characterization of the Ampere GPU's instruction-set architecture (ISA).

\definecolor{backcolour}{rgb}{0.95,0.95,0.92}
% \definecolor{codepurple}{rgb}{0.58,0,0.82}
\definecolor{codepurple}{rgb}{1,0,0}
\definecolor{codegreen}{rgb}{0,0.6,0}
\definecolor{codegray}{rgb}{0.5,0.5,0.5}

\lstdefinelanguage{PTX}
{
  morekeywords={ld, st, mov, add, sub, setp, cvta},
  ndkeywords={global, clock64, clock, param, ret, shared, bra, reg}
}

\begin{figure*}[t!]
    \centering
    \begin{minipage}{.5\textwidth}
        \centering
           \begin{lstlisting}[language=PTX,
                      numbers=left, 
                      xleftmargin=0.4in, 
                      xrightmargin=0.4in, 
                      basicstyle=\scriptsize,
                      frame=single, 
                      backgroundcolor=\color{backcolour},
                      numberstyle=\scriptsize\color{codegray},
                      keywordstyle=\color{codepurple},
                      ndkeywordstyle=\color{blue}
                     ]
.visible .entry _Z3AddPi(
        .param .u64 _Z3AddPi_param_0
)
{
    .reg .b32   %r<100>;
    .reg .b64   %rd<100>;

    ld.param.u64    %rd1, [_Z3AddPi_param_0];
    cvta.to.global.u64  %rd4, %rd1;

    add.s32         %r5, 5, %r3;
    add.s32         %r7, %r5, 2;
    mov.u32         %r1, %clock;
    add.u32         %r11, 6, %r7;
    add.u32         %r12, %r5, 7;
    add.u32         %r13, %r12, %r1;
    mov.u32         %r2, %clock;
    sub.s32         %r8, %r2, %r1;

    st.global.u32   [%rd4], %r8;
    st.global.u32   [%rd4 + 8], %r11;
    st.global.u32   [%rd4 + 16], %r12;
    st.global.u32   [%rd4 + 20], %r13;
    ret;
}
\end{lstlisting} 
\caption{Computing unsigned add instruction latency.}
        \label{fig:PTX_ALU}
    \end{minipage}%
    \begin{minipage}{0.5\textwidth}
        \centering
        \begin{lstlisting}[language=PTX,
                    numbers=left, 
                    xleftmargin=0.4in, 
                    xrightmargin=0.4in, 
                    basicstyle=\scriptsize,
                    frame=single, 
                    backgroundcolor=\color{backcolour},
                    numberstyle=\scriptsize\color{codegray},
                    keywordstyle=\color{codepurple},
                    ndkeywordstyle=\color{blue}
                 ]
        mov.u64             %r19,%rd4;
        mov.u64             %r40,0;
$Mem_store:
        st.wt.global.u64   [%r19], %r19+8;
        st.wt.global.u64   [%r19+8], %r19+16;
        st.wt.global.u64   [%r19+16], %r19+24;
        st.wt.global.u64   [%r19+24], %r19+32;
        add.u64            %r19,%r19,32;
        add.u64            %r40,%r40,32;
        setp.lt.u64 %p1, %r40, 52268760; 
        @%p1 bra           $Mem_store;

        mov.u64                 %r40,0;
        mov.u64	                %r1, %clock64;
$Mem_load:
        ld.global.cv.u64 	%r4, [%rd4 ];
	ld.global.cv.u64 	%r16, [%r4];
        ld.global.cv.u64 	%r17, [%r16];
        ld.global.cv.u64 	%r20, [%r17];
        add.u64                 %r40,%r40,32;
        setp.lt.u64 %p1,        %r40, 262144; 
        @%p1 bra $Mem_load;

        mov.u64 	%r2, %clock64;
        sub.s64         %r7, %r2, %r1;
\end{lstlisting}
% \vspace{-1ex}
        \caption{Computing L2 cache and global memory access latency}
% \vspace{-2ex}
         \label{fig:PTX_mem_glo}
    \end{minipage}
\vspace{-1ex}
\end{figure*}

\begin{figure}
\begin{lstlisting}[language=PTX,
            numbers=left, 
            xleftmargin=0.5in, 
            xrightmargin=0.5in, 
            basicstyle=\scriptsize,
            frame=single, 
            backgroundcolor=\color{backcolour},
            numberstyle=\scriptsize\color{codegray},
            keywordstyle=\color{codepurple},
            ndkeywordstyle=\color{blue}
         ]
    //reading from shared memory
    mov.u64	         %r1, %clock64;
    ld.shared.u64        %r25,[shMem1];
    add.u64              %r40,%r25,32;
    mov.u64 	         %r2, %clock64;

    sub.s64              %r7, %r2, %r1;
    add.u64		 %r22, %r7, 10;

    //storing to the shared memory 
    mov.u64	         %r1, %clock64;
    st.shared.u64        [shMem1],50;
    add.u64              %r24,%r23,32;
    mov.u64 	         %r2, %clock64;

    sub.s64              %r16, %r2, %r1;
    add.u64		 %r22, %r16, 10;
\end{lstlisting}
\vspace{-1ex}
    \caption{Computing device shared memory access latency.}
\vspace{-2ex}
     \label{fig:PTX_shmem}
\end{figure}

%====================================================
%====================================================
\section{Related work}
\label{sec:Related_work}

Various work have been conducted to dissect every undisclosed microarchitecture characteristic of the GPU~\cite{arafa2019low,arafa2020verified,mei2016dissecting,jia2018dissecting,jia2019dissecting,sun2022dissecting} using microbenchmarks.  Unlike the Nvidia Ampere architecture, the older architectures such as Kepler, Fermi, Volta, and Turing are heavily studied in the literature. Some focused on the instruction level~\cite{wong2010demystifying,arafa2019low}, while others focused on the hardware unit itself~\cite{mei2016dissecting, markidis2018nvidia,fasi2021numerical}. In this section, we present some of these works in more detail.

Wong~\etal~\cite{wong2010demystifying} was the first to introduce microbenchmarks to measure the latency and the throughput of different types of memories and instructions. In \cite{van2022isolating}, the authors modified some microbenchmarks to isolate the GPU features and study each separately. They studied the effect of the number of warps and blocks on the throughput for the memory, arithmetic, and tensor cores operations. They calculated the latencies per block, not per instruction. Other works~\cite{bombieri2016mipp,jia2018dissecting,jia2019dissecting} calculated instructions and memory throughput and latencies for the Kepler, Volta, and Turing architectures. However, \cite{bombieri2016mipp} added more details about energy consumption. In the same spirit, Mei~\etal~\cite{mei2016dissecting} presented a microbenchmark for calculating the throughput and latencies of different types of memory units on older architectures such as Fermi, Kepler, and Maxwell.

Other researchers focused on profiling the tensor core in Volta and Turing architectures~\cite{markidis2018nvidia,raihan2019modeling,yan2020demystifying}. Recently, Sun~\etal~\cite{sun2022dissecting} tried to dissect the tensor core in the Ampere architecture. The authors focused on investigating the matrix multiply-accumulate using the \textit{MMA} API, which gets executed on the tensor cores. They did not provide results for the \textit{WMMA} API. In~\cite{kothiya2014understanding}, the authors demonstrated the mapping of the PTX to the SASS instructions for the tensor core operations. Fasi~\etal~\cite{fasi2021numerical} developed a microbenchmark to investigate the tensor core numerical behavior and proved that the tensor core supports the subnormal number.

While all the previous work presents good progress, none focused on the clock cycles latency for all data types of each instruction while demonstrating the PTX and SASS mapping for each instruction. Moreover, to the best of our knowledge, we are the first to investigate every \textit{WMMA} Tensor Core instruction clock cycles and throughput with different data types for the Nvidia Ampere architecture. Finally, Our work can be easily extended for future architectures.

%=======================================================
%=======================================================
\section{Methodology}
\label{sec:Methodology}

In this section, we introduce the microbenchmark design details. Our work is based on extending the microbenchmark presented in~\cite{arafa2019low} to calculate the clock cycles per instruction for the Nvidia Ampere (AI100) GPU. We modified the microbenchmark to calculate the latency for dependent and independent instructions. Furthermore, we extended the code to calculate the clock cycles latency for the different types of memory units and the tensor core instruction.

The microbenchmarks are directly written in PTX, a pseudo assembly intermediate and architecture-independent  ISA across all Nvidia. However, writing directly in PTX ISA can be tricky since the compiler translates the PTX code into another architecture-dependent ISA, SASS. There is not much information available on how the compiler does the mapping from PTX to SASS. For instance, the compiler can optimize multiple PTX instructions into one SASS instruction. In order to overcome these limitations and ensure the proper instructions get executed are the ones we need, we dynamically read the SASS instruction trace at the run time of each PTX microbenchmark written. We use the \textit{Tracing Tool} from \textit{PPT-GPU}~\cite{arafa2021hybrid} to do that. We then tweak the PTX microbenchmark by trial and error to give us the correct SASS results.

\subsection{Instructions Clock Cycles Latency}
\label{subsec:inst-lat}

We used only one thread per block to measure the instruction latency. We have two steps. \textit{First}, We run a code that calculates the clock cycles for the studied instruction with a specified data type. For instance, the code shown in Figure~\ref{fig:PTX_ALU} calculates the latency of the add instruction where the operands are 32-bit registers. In general, measuring the latency can be performed by reading the clock before and after the instruction, as shown in Figure \ref{fig:PTX_ALU} lines 13 and 17. Then, we subtract the two clock readings (line 18) to calculate the difference or the required latency. We execute the three independent add instructions (lines 14-16). We also used dependent instructions and found that latency increased compared to independent instructions. Finally, we return the latency value to the main CUDA function and divide it by 3 to calculate the number of cycles for each instruction. We use 3 instructions to overcome the first launch overhead. We found that executing only one instruction will result in an unexpected higher number of cycles. Table~\ref{table:tabel_CPI} shows an example for \textit{add.u32} instruction. The first instruction takes around 5 cycles. Nevertheless, when we use more than 3 instructions, the average number of cycles per instruction (CPI) is 2.

\renewcommand{\arraystretch}{1.2}
\begin{table}[t!]
\caption{The relation between the number of instructions and the average cycles for add.u32 instruction}
\begin{center}
\begin{tabular}{|c|c|}
\hline
\cline{2-2} 
\textbf{\# instrs} & \textbf{\textit{CPI}} \\
\hline
1 & 5  \\
2 & 3  \\
3 & 2  \\
4 & 2 \\
\hline
\end{tabular}
% \vspace{-5ex}
\label{table:tabel_CPI}
\end{center}
\end{table}

\begin{figure}[t!]
\centering
\subfigure[Using 32 bit clocks register]{%
\label{fig:clocks32}%
\includegraphics[width=\linewidth]{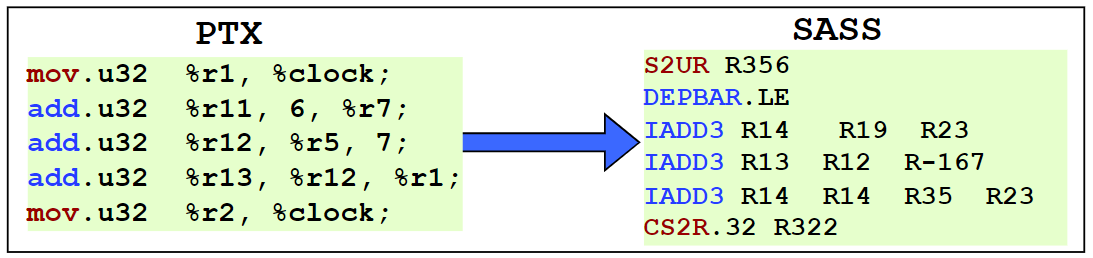}}%
\qquad
\subfigure[SASS 64 bit clocks register]{%
\label{fig:clocks64}%
\includegraphics[width=\linewidth]{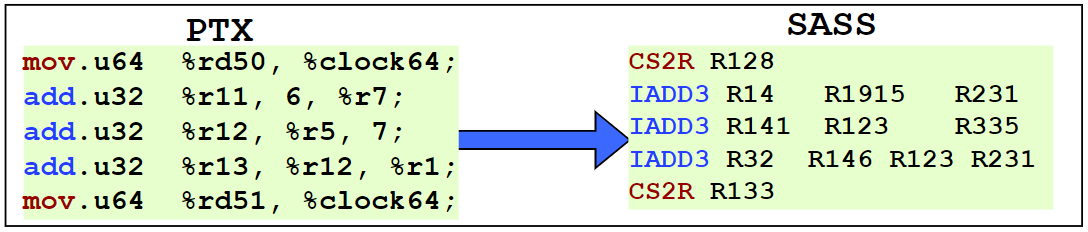}}%
\caption{Mapping of PTX to SASS when using 32 and 64 bits clock registers}
\label{fig:clck_regst_example}
% \vspace{-3ex}
\end{figure}

\textit{Second}, we inspect the sass instruction using the dynamic \textit{Tracing Tool} from \textit{PPT-GPU}~\cite{arafa2019ppt} to ensure that the mapping from PTX to SASS is correct and no additional overhead or instruction is added at runtime by the compiler. The PTX code shown in Figure \ref{fig:clocks32} provides an inaccurate latency for the add instruction when storing the clock in 32-bit registers. The dynamic SASS instruction trace shows a barrier between the two clock readings, as shown in the second instruction of the SASS part. This barrier causes a considerable change in the results (around ~33 cycles increase in this case). One method to overcome this barrier is to use the 64-bit registers to store the clocks, which remove the barrier and provide an accurate measurement, as shown in Figure \ref{fig:clocks64}. The CPI for the first and second cases are 13 and 2 cycles, respectively. Finally, we calculate the clock overhead using two consecutive clock reading instructions and find that it equals 2 cycles.

\subsection{Memory Units Access Latency}
\label{subsec:mem-latency}

To calculate \textbf{global, L2, and L1 cache memories latency}, we use a pointer chasing technique, in which each array elements are dependent on the previous ones. This technique forces the reading operations to be serialized to calculate the latency correctly. Otherwise, many reading operations can be issued simultaneously, which makes the latency measurements inaccurate. Figure~\ref{fig:PTX_mem_glo} shows the PTX microbenchmark for the memory latency calculations. Line 1 moves the array address to the \%r19 register. Then, we start a counter with a zero value in the \%r40 register. This counter is used to iterate over the array of elements. Lines 3 and 9-11 represent the loop instructions. Lines 4 to 7 are used to store the array of elements in which each element is dependent on the previous one. After storing the results, we use the instructions from lines 14 to 24 to read the clocks while reading every element in the array. From 16 to 19, we have 4 load instructions to load 4 values, which are repeated to read all the array elements. 

The \textit{ld} instruction can be used with many operators such as \textit{cv, ca, and cg}. Each operator has its usage. \textit{ca} is used to cache on all available levels (global-L1-L2) while the \textit{cg} caches only in L2. On the contrary, we use \textit{cv} because it bypasses the caches, which we need when we calculate the global memory latency. We use 4 instructions because we found that in many cases, the compiler unrolls the loops by 4 when we inspected the dynamic trace of some Cuda applications that use loops. The difference between the global memory code and the l2 cache code is the operator used with the \textit{ld} instruction and the number of the elements in the array. For the L2 cache, we use the \textit{cg} operator, and the total size of the array elements must be less than the L2 size, while for the global memory code, it must be larger than the L2 cache to avoid L2 cache residency. Likewise, we repeat the same methodology with the \textit{ca} operator to calculate  \textbf{the L1 cache latency}.

For the \textbf{shared memory}, we load and store instructions between reading the clocks, as shown in lines 3-12 of Figure \ref{fig:PTX_shmem}. However, we needed to add another instruction that depends on the \textit{ld} or \textit{st} instructions to prevent the compiler from executing the clock reading instruction before finishing, as shown in lines 4-13.

\definecolor{backcolour}{rgb}{0.95,0.95,0.92}
%\definecolor{codepurple}{rgb}{0.58,0,0.82}
\definecolor{codepurple}{rgb}{1,0,0}
\definecolor{codegreen}{rgb}{0,0.6,0}
\definecolor{codegray}{rgb}{0.5,0.5,0.5}

\lstdefinelanguage{PTX}
{
  morekeywords={ld, st, mov, sub,wmma},
  ndkeywords={global, clock, param, ret,load_matrix_sync,mma_sync,store_matrix_sync,unsigned,int,if,printf}
}

\begin{figure}
   \begin{lstlisting}[language=PTX,
              numbers=left, 
              xleftmargin=0.2in, 
              xrightmargin=0.1in, 
              basicstyle=\fontsize{5.5}{7}\selectfont\ttfamily,
              frame=single, 
              backgroundcolor=\color{backcolour},
              numberstyle=\scriptsize\color{codegray},
              keywordstyle=\color{codepurple},
              ndkeywordstyle=\color{blue}
             ]
__global__void wmma_example(atype *a, btype *b, ctype *c,dtype *d)
{
   unsigned int start,start_time=0,end_time=0;
   // Part 1: Declare the fragments
   wmma::fragment<wmma::matrix_a, M, N, K, atype , LAYOUT_A> a_frag;
   wmma::fragment<wmma::matrix_b, M, N, K, btype , LAYOUT_B> b_frag;
   wmma::fragment<wmma::accumulator, M, N, K, ctype> c_frag;

   // Part 2: loading the values from the memory
   wmma::load_matrix_sync(a_frag, a, A_STRIDE);
   wmma::load_matrix_sync(b_frag, b, B_STRIDE);
   wmma::load_matrix_sync(c_frag, c, C_STRIDE,LAYOUT_C);

    // Part 3: running the multiple+add on matrices 
   start_time=clock();
   for (int i=0; i<iters*iters1; i++){
      wmma::mma_sync(c_frag, a_frag, b_frag, c_frag);
      wmma::mma_sync(c1_frag, a1_frag, b1_frag, c1_frag);
      wmma::mma_sync(c2_frag, a2_frag, b2_frag, c2_frag);
      wmma::mma_sync(c3_frag, a3_frag, b3_frag, c3_frag);
   }
   end_time=clock();
   
   // Part 4: store the values from the memory
   wmma::store_matrix_sync(d, c_frag, C_STRIDE, LAYOUT_C);

   if(threadIdx.x==1)
     printf("CLOCK for all=%d \t %d \n",((end_time-start_time)-2)
     /(4*iters1),tid); 
}
\end{lstlisting} 
% \vspace{-1ex}
\caption{Computing the tensor core WMMA instruction clock cycles latency for U8 data type.}
% \vspace{-1ex}
\label{fig:TC_latency}
\end{figure}

\begin{figure}[t!]
      \centering
      \includegraphics[width=.6\linewidth]{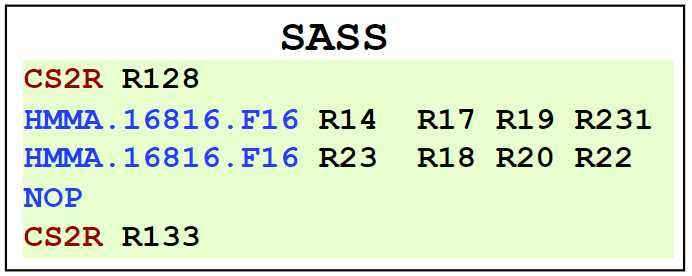}
      \caption{SASS instructions of running one TC instruction}
      \label{fig:SASS_TC}
    %   \vspace{-1ex}
\end{figure}

\subsection{Tensor core Instructions Latency and Throughput}

The Tensor core (TC) unit is a very trivial unit for accelerating machine learning applications. Each SM in the Ampere architecture contains 4 tensor cores that can run the multiple-add operation on 3 matrices in the form \textit{D=A*B+C} where A, B are the inputs and C, D are the accumulators. Unlike the Volta, which supports only \textit{fp16} precision for the inputs, the Ampere architecture supports many types, such as \textit{FP16, bf16, tf32, f64, u8, u4 and the single bit}. The general arithmetic instructions use one thread to execute and can communicate only through the global or shared memory. On the other hand, the TC instructions use all the 32 instructions in the dedicated warp. To demystify the Ampere architecture's TC instructions with the new data types, we designed a special microbenchmark is written in Cuda programming language. The microbenchmarks are inspired by Jia~\etal~\cite{jia2018dissecting} work, which focused on Volta architecture.

Some of the new data types that were introduced in the Ampere architecture are still in the experimental stage, as mentioned in PTX and CUDA documentation~\cite{Nvidia_Cuda}. Moreover, because each data type has its shapes, stride, and layout, we use a different function to calculate the latencies of each type. Figure~\ref{fig:TC_latency} shows the code used to calculate the TC instruction latency of \textit{U8} data type. Lines 5 to 7 create a fragment in which the registers are prepared to get the matrix elements to be stored. We create 4 fragments; however, we do not write all to make the shape smaller. Then, we load the data from memory (lines 10-12), and the same goes for the other fragments. As previously explained, we read the clocks before and after the TC WMMA execution (lines 15 to 22) and subtract them before printing, as shown in lines 28-29. From lines 16 to 21, we run 4 TC instructions (1 per TC) numerous times. We used 4 instructions because we found that calculating the latency from one TC with one instruction provides inaccurate measurement. For example, Figure \ref{fig:SASS_TC} shows the dynamic SASS instructions of running one instruction on one TC. The NOP instructions refer to a warp synchronization in PTX, and we found that the latency here is not the same as mentioned in the white paper. This also happens when we run one instruction several times. Finally, we calculate the latency per instruction and print them by lines 28-29. A similar method is used to calculate the TC's WMMA throughput.

%========================================================
%========================================================
\section{Results}
\label{sec:Results}

In this section, we present the detailed setup and results. We first show the instructions clock cycles latency. Next, we explain the memory access latencies. Finally, We present the Tensor Core latency and throughput. We run all the microbenchmarks on the Nvidia \textit{\textbf{Tesla AI100}} GPU. 

\subsection{Instructions Latency}

We found that dependency directly affects the instructions clock cycle latency. Hence, we rerun the microbenchmark with a sequence of dependent instructions (shown in Figure~\ref{fig:PTX_ALU}), replacing the dependent sequence with another sequence of independent instructions. Table~\ref{table:tabel_dep_no_dep} shows the CPI for dependent and independent sequences for some of the instructions. For instance, single precision add instruction shows 4 and 2 cycles, respectively. We also found that with no dependency, the 3 \textit{add.u32} instructions mentioned in Figure \ref{fig:PTX_ALU} are mapped to the same sass instruction (\textit{IADD}) as shown in Figure \ref{fig:clocks32}. Nevertheless, PTX instructions may be converted to different instructions when we use three dependent instructions. For example, the \textit{add.u32} PTX instruction can be mapped to \textit{IADD3} or \textit{IMAD.IADD} with the dependency case.

Table~\ref{table:ampere_cuda_11} depicts the various PTX-SASS instructions with their measured clock cycles latencies. We have a separate PTX kernel (microbenchmark) for each field in the table. 

Next, we discuss some additional insights we found while generating the results:

\textbf{1:} The \textit{mad} instruction runs on the floating pipeline, not the integer pipeline, even if we use it with integer values. This can be proven by the following:
\begin{itemize}
\item The PTX instruction\textit{mad.lo.u32} in Table~\ref{table:ampere_cuda_11} is mapped to the SASS \textit{FFMA} (floating multiple-add). 
\item We created a special code that runs two add instructions and two mad instructions using one thread, and we found that the total number of cycles is around 4 cycles. This means that each one of the four instructions takes 1 cycle. It proves that each of the two types is executed simultaneously on different pipelines.
\end{itemize}

Showing that \textit{mad} instruction uses another pipeline explains why the dependent PTX \textit{add.u23} instruction is mapped to the SASS (\textit{IMAD.IADD}) instruction in some cases. The compiler is trying to use the floating pipeline while waiting for the integer pipeline to commit.

\textbf{2:} Except for \textit{bfind}, \textit{min} and \textit{max} instructions there is no difference in clock cycles or mapping between PTX to SASS when using a signed or an unsigned instruction. For instance, \textit{add.u64} and \textit{add.s64} provide the same mapping and the same latency.

\textbf{3:} Usually, a \textit{mov} or \textit{add} instruction is used to initialize a register with a value before using this register as an input operand to the instruction that we need to calculate its latency. However, in some cases, we found that the clock cycles and the PTX-SASS mapping change depending on how the inputs are initialized. For example, the PTX \textit{neg.f32} is mapped to the SASS \textit{FADD} when we use add instruction to initialize the inputs. on the other hand, it merges the \textit{mov} and the \textit{neg} instructions together in one SASS instruction (\textit{IMAD.MOV.u32}) when we use \textit{mov} for initializing. The same happens for the \textit{abs.f32} instruction.

\renewcommand{\arraystretch}{1.2}
\begin{table}[t!]
\caption{The CPI for dependent and independent instructions}
\begin{center}
\begin{tabular}{|c|c|c|}
\hline
\cline{3-3} 
\textbf{\# instrs} & \textbf{\textit{CPI for dependent}}  & \textbf{\textit{CPI for independent}} \\
\hline
add.f16 & 3 & 2  \\
add.u32 & 4 & 2  \\
add.f64  & 5 & 4  \\
mul.lo.u32 & 3 & 2  \\
mad.rn.f32 & 4 & 2  \\
\hline
\end{tabular}
% \vspace{-4ex}
\label{table:tabel_dep_no_dep}
\end{center}
\end{table}

%====================================
\renewcommand{\arraystretch}{1.2}
\begin{table*}[t!]\scriptsize
\caption{The Tensor cores latencies and throughput}
\begin{center}
\noindent\makebox[\textwidth]{\begin{tabular}{|c||c||c||c||c||c|}
\hline
\cline{2-5} 
\textbf{ Supported shapes} & \textbf{Inputs} & \textbf{Accumulators} & \textbf{Cycles} & \textbf{Measured-theoretical } & \textbf{Instructions}\\
\hline
m16n16k16 - m8n32k16 - m32n8k16  & .f16 &.f16 &16 &311-312 GB/s
&\multirow{2}{*}{} \textbf{PTX}: \textcolor{blue}{wmma.mma.sync.aligned.row.row.m16n16k16.f16.f16} \\&&&&&
 \textbf{SASS}: \textcolor{red}{2*HMMA.16816.F16} {– each inst. is 8 cycles}
  \\
    \hline
m16n16k16 - m8n32k16 - m32n8k16  & .f16 &.f32 &16 &310-312 GB/s
&\multirow{2}{*}{}\textbf{PTX}: \textcolor{blue}{wmma.mma.sync.aligned.row.row.m16n16k16.f16.f32}
 \\&&&&&\textbf{SASS}: \textcolor{red}{2*HMMA.16816.F32} {– each inst. is 8 cycles }
  \\
  \hline

m16n16k16 - m8n32k16 - m32n8k16  & .bf16 &.f32 &16&310-312 GB/s
&\multirow{2}{*}{}\textbf{PTX}: \textcolor{blue}{wmma.mma.sync.aligned.row.row.m16n16k16.f32.bf16.bf16.f32}

 \\&&&&& \textbf{SASS}: \textcolor{red}{2*HMMA.16816.F32.BF16} {– each inst. is 8 cycles }
  \\
  \hline

m16n16k8  & .tf32 &.f32 &16 &132-156 GB/s
&\multirow{2}{*}{}\textbf{PTX}: \textcolor{blue}{wmma.mma.sync.aligned.row.row.m16n16k8.f32.tf32.tf32.f32}
 \\&&&&& \textbf{SASS}: \textcolor{red}{4*HMMA.1684.F32.TF32} {– each inst. is 4 cycles  }
  \\
  \hline

m8n8k4   & .f64 &.f64 &16&19-19.5 GB/s
&\multirow{2}{*}{}\textbf{PTX}: \textcolor{blue}{wmma.mma.sync.aligned.row.row.m8n8k4.f64.f64.f64.f64.rn}
 \\&&&&& \textbf{SASS}: \textcolor{red}{1*DMMA.884} {– each inst. is 16 cycles   }
  \\
  \hline
 
m16n16k16 - m32n8k16 - m8n32k16  & .u8 &.u32 &8&594-624 GB/s
&\multirow{2}{*}{}\textbf{PTX}: \textcolor{blue}{wmma.mma.sync.aligned.row.row.m16n16k16.s32.u8.u8.s32}
 \\&&&&& \textbf{SASS}: \textcolor{red}{2*IMMA.16816.U8.U8}  {– each inst. is 4 cycles}
  \\
  \hline

m8n8k32  & .u4 &.u32 &4&1229-1248 GB/s
&\multirow{2}{*}{}\textbf{PTX}: \textcolor{blue}{wmma.mma.sync.aligned.row.col.m8n8k32.s32.u4.u4.s32}
 \\&&&&& \textbf{SASS}: \textcolor{red}{1*IMMA.8832.U4.U4}  {each inst. is 4 cycles}
  \\
  \hline

\end{tabular}}
\label{tabel:TC_latencies}
\end{center}
% \vspace{-4ex}
\end{table*}
%====================================

\textbf{4:} Although many PTX instructions have a 1-to-1 mapping to SASS, others such as \textit{div, rem, sinf, and cosf} are translated to multiple different SASS instructions.

\textbf{5:} Not all instructions with the same data type have the same latency. More specifically, \textit{mad.lo.u64} is mapped to an \textit{IMAD} SASS instruction which takes only 2 cycles. However, the double precision add, sub and fma instructions take 4 cycles each.

\textbf{6:} For the \textit{testp} instruction, the latency depends on the state.

\subsection{Memory Access Latencies}
The observed latencies of the different types of memories are shown in Table~\ref{table:tab1_mem}. The global memory latency is around 290 cycles. This value does not include the cache misses latencies because we prevent caching at all levels. This number is improved compared to Turing architecture which is 434 cycles~\cite{arafa2019low}. The L2 access latency is 200 cycles compared to 188 cycles for Turing architecture. Furthermore, the L1 cache hit for both Ampere and Truing architectures is 33 and 32 cycles, respectively. For the shared memory, we found that the store access latency takes a smaller value than the load instruction, 23 and 19 for load and store, respectively. 

\subsection{Tensor Core Latencies and Throughput}

The Ampere architecture ISA provides various SASS instructions that run on the Tensor Core, which supports the newly added data types. The Volta Architecture's ISA has only the \textit{HMMA.884} SASS instruction handles all Tensor Core operations (single and mixed-precision operations). For Turing, two kinds of the \textbf{HMMA} SASS instructions exist which runs on different input shapes, \textit{HMMA.1688} and \textit{HMMA.884}~\cite{yan2020demystifying}.

Table~\ref{tabel:TC_latencies} depicts the Ampere architecture's TC instructions. More specifically, \textit{DMMA.884}, \textit{IMMA.16816} and \textit{IMMA.8832} were added to handle the FP64, U8 and U4 data types, respectively. Each PTX instruction of each data type is translated to a different number of SASS instructions. For the FP16 BF16 and U8 inputs, the PTX is translated to 2 instructions. The TensorFloat-32 (TF32) precision is mapped to 4 SASS instructions, while the FP64 and U4 are mapped to only 1 instruction. These differences are related to the difference between the supported PTX shapes and the shapes that the SASS can work on it. For example, in Table~\ref{tabel:TC_latencies}'s first row, the PTX instruction can use many shapes such as 16×16×16, but the SASS can only work on 16×8×16. So, 2 SASS instructions are needed to iterate over the PTX shape. However, the physical TC implementation can perform 8*4*8~\cite{sun2022dissecting}. While it is previously mentioned in~\cite{raihan2019modeling} that the TC latencies are shape-dependent for Turing, we found that different shapes for the same data type do not affect the calculated latency. It can vary from one type to another in Ampere architecture. Our observations for the TC throughput and latencies shown in the table are consistent with the behavior mentioned in the white paper~\cite{Nvidia_ampere_arch_doc}. Finally, We noticed that for all half floating precision (fp16 and bf16) inputs, SASS instruction \textit{MOVM.16.MT88} is used for loading a matrix to the TC. In general, the \textit{MOVM} SASS instruction is used to move a matrix with a transpose. The number of issued \textit{MOVM} instructions depends on the matrix shape and the layout (row or column major). For example, if we used A and B matrices as row major in the PTX code, then the \textit{MOVM} instructions are used to transpose the B matrix to multiply each row from A with each column from B. However, when we use both as Column major, the \textit{MOVM} instruction is used with the A and C matrices. It transposes A and C before execution and transposes C after the execution. Finally, if A is a row-major and B is a column-major, the MOVM instruction does not exist in the trace.
We used the same way motioned above for the latency calculations to calculate the memories throughput. The observations are quite similar to the throughput values mentioned in the white paper. 

\renewcommand{\arraystretch}{1.2}
\begin{table}[ht!]
\caption{The memory accesses latencies}
\begin{center}
\begin{tabular}{|c|c|}
\hline
\cline{2-2} 
\textbf{Memory type} & \textbf{\textit{CPI (cycles)}} \\
\hline
Global memory & 290  \\
L2 cache & ~200  \\
L1 cache & 33 \\
Shared Memory (ld/st) & (23/19)  \\
\hline
\end{tabular}
\label{table:tab1_mem}
\end{center}
% \vspace{-3ex}
\end{table}

\section{Conclusion}

This paper demystifies the instructions, memories, and tensor cores for Nvidia Ampere architecture. We perform a detailed analysis of the PTX instructions latency while showing their SASS translation. The presented microbenchmarks are portable and can be extended for future architectures. In addition, we pointed out the microarchitecture instructions of the tensor cores and their latencies for all data types supported by the Ampere architecture. Finally, we calculate the memory latency while building the pointer chasing method for both global memory and L2 cache. This work can help in understating the hardware from the microarchitecture point of view, leading to better-optimized applications and workloads.

%====================================
% \input{lat-table}
\renewcommand{\arraystretch}{1.1}
\begin{table*}[ht!]\scriptsize
  \centering
  \caption{Instructions clock cycles for the (\textit{Amepere A100}) GPU}
% \vspace{-2ex}
  \noindent\makebox[\textwidth]{\begin{tabular}{|c|c|c|c|c|c|}
  \hline
  \textbf{PTX} & \textbf{SASS} & \textbf{cycles} & \textbf{PTX} & \textbf{SASS} & \textbf{cycles} \\
  \hline
  \multicolumn{3}{|c|}{\textbf{Add / sub instruction}} &\multicolumn{3}{|c|}{\textbf{Min/Max instructions}}\\
 \hline
   add.u16 & UIADD3   & 2 &Min.u16&ULOP3.LUT+UISETP.LT.U32.AND+USEL&8 \\
  \hline
   addc.u32& IADD3.X & 2 & min.u32&IMNMX.U32 
&2\\
  \hline
   add.u32 & IADD & 2 &min.u64
&UISETP.LT.U32.AND+2*USEL & 8 \\
  \hline
  add.u64 & UIADD3.x+ UIADD3& 4 &min.s16&PRMT+IMNMX&4 \\
  \hline
  add.s64 & UIADD3.x+UIADD3 & 4 &min.s32&IMNMX
&2 \\
  \hline
  add.f16 & HADD     & 2 &Min.s64
&UISETP.LT.U32.AND+UISETP.LT.AND.EX+2*USEL&8 \\
  \hline
  add.f32& FADD& 2&min.f16 &HMNMX2+PRMT&4  \\
  \hline
  add.f64& DADD& 4 &min.f32 &FMNMX&2\\
  \hline
  \multicolumn{3}{|c|}{\textbf{Mul instruction}} &min.f364
&DSETP.MIN.AND+IMAD.MOV.U32+UMOV+FSEL
&10\\
 \hline
   mul.wide.u16       & LOP3.LUT+IMAD    & 4 & \multicolumn{3}{|c|}{\textbf{Neg instruction}}\\
  \hline
   mul.wide.u32   & IMAD    & 4  &neg.s16&UIADD3+UPRMT&5\\
  \hline
   mul.lo.u16 & LOP3.LUT+IMAD & 4 &neg.s32&IADD3&2 \\
  \hline
 mul.lo.u32& IMAD& 2 &neg.s64&IMAD.MOV.U32+HFMA2.MMA+MOV+UIADD3
&~10 \\
  \hline
 mul.lo.u64& IMAD& 2 &neg.f32&FADD or IMAD.MOV.U32 *&2 \\
  \hline
  mul24.lo.u32& PRMT + IMAD & 3 &neg.f64&DADD+(UMOV)&4\\ 
  \hline
  mul24.hi.u32& UPRMT+USHF.R.U32.HI+IMAD.U32+PRMT
 & 9 & \multicolumn{3}{|c|}{\textbf{FMA instruction}} \\
  \hline
  mul.rn.f16& HMUL2& 2 &fma.rn.f16&HFMA2&2\\
  \hline
  mul.rn.f32 & FMUL & 2&fma.rn.f32&FFMA&2\\
  \hline
  mul.rn.f64 & DMUL & 4 &fma.rn.f64&DFMA&4\\
  \hline
  \multicolumn{3}{|c|}{\textbf{MAD Instruction}} &\multicolumn{3}{|c|}{\textbf{Sqrt Instruction}}\\
  \hline
   mad.lo.u16       & LOP3.LUT+IMAD   & 4&sqrt.rn.f32
&[multiple instrs including MUFU.RSQ]&190-235
 \\
  \hline
  mad.lo.u32 & FFMA  & 2& sqrt.approx.f32&[multiple instrs including MUFU.SQRT]&2-18
\\
  \hline
   mad.lo.u64    & IMAD  & 2&sqrt.rn.f64&[multiple insts including MUFU.RSQ64] &260-340 \\
  \hline
   mad24.lo.u32   & SGXT.U32 + IMAD   & 4 &\multicolumn{3}{|c|}{\textbf{Rsqrt Instruction}}\\
  \hline
  mad24.hi.u32  & USHF.R.U32.HI+UIMAD.WIDE.U32+2*UPRMT+IADD3
  & 11&rsqrt.approx.f32&[multiple insts including MUFU.RSQ]&2-18 \\
  \hline
   mad.rn.f32& FFMA& 2&rsqrt.approx.f64&MUFU.RSQ64H&8-11\\
  \hline  
  mad.rn.f64& DFMA& 4 &\multicolumn{3}{|c|}{\textbf{Rcp Instruction}}\\
  \hline  
  
  \multicolumn{3}{|c|}{\textbf{Sad Instruction}} &rcp.rn.f32
&[multiple insts including MUFU.RCP]&198\\
  \hline
   sad.u16/s16 & (2*LOP3) +ULOP3+ VABSDIFF   & 6&rcp.approx.f32&[multiple insts including MUFU.RCP]&23 \\
  \hline
  sad.u32/s32 & VABSDIFF +IMAD (1 IMAD + 1 Umov for 3 instrs)  & 3&rcp.rn.f64
&[multiple insts including MUFU.RCP64H] &244 \\
  \hline
   sad.u64/s64& UISETP.GE.U32.AND+UIADD+IADD  & 10&ex2.approx.f32
 &FSTEP + FMUL + MUFU.EX2 + FMUL&14\\
  \hline

  \multicolumn{3}{|c|}{\textbf{Div / Rem Instruction}}&\multicolumn{3}{|c|}{\textbf{Pop Instruction}} \\
  \hline
   rem/div.u16/s16  &  multiple instructions   & ~290&popc.b32S&POPC
&6 \\
  \hline
 rem/div.s32/u32   &  multiple instructions  & 66&popc.b64&2*UPOPC + UIADD3&7 \\
  \hline
  rem/div.u64/s64&   multiple  instructions & ~ 420 &\multicolumn{3}{|c|}{\textbf{Clz
 Instruction}}\\
  \hline
 div.rn.f32  &  multiple instructions &   ~ 525 &clz.b32&FLO.U32 + IADD&7\\
  \hline
   div.rn.f64  & multiple  instructions&  ~ 426 &clz.b64&UISETP.NE.U32.AND+USEL+UFLO.U32+2*UIADD3&13\\
  \hline

  \multicolumn{3}{|c|}{\textbf{Abs Instruction}}&\multicolumn{3}{|c|}{\textbf{Bfind
 Instruction}} \\
  \hline
   abs.s16   &  PRMT+IABS+PRMT    & 4&bfind.u32
&FLO.U32&6 \\
  \hline
 abs.s32    &  IABS   & 2&bfind.u64&FLO.U32+ISETP.NE.U32.AND+IADD3+BRA
&164\\
  \hline
  abs.s64 & UISETP.LT.AND+UIADD3.X +UIADD3+2*USEL  & ~ 11&bfind.s32&FLO&6  \\
  \hline
 abs.f16     &  PRMT   &   1 &bfind.s64&multiple instructions&195\\
  \hline
   abs.ftz.f32 &  FADD.FTZ  &  2 &\multicolumn{3}{|c|}{\textbf{testp  Instruction}} \\
  \hline  
   abs.f64&  DADD or (DADD+UMOV)   & 4 & testp.normal.f32 
&IMAD.MOV.U32+2*ISETP.GE.U32.AND & 0 or 6  \\
  \hline

\multicolumn{3}{|c|}{\textbf{Brev  Instruction}}
&testp.subnor.f32 &ISETP.LT.U32.AND &0 or 6\\
  \hline
   brev.b32    &  BREV + SGXT.U32     & 2& testp.normal.f64 
&2*UISETP.LE.U32.AND+2*UISETP.GE.U32.AND
&13\\
  \hline
   brev.b64    &  2*UBREV+MOV     & 6&testp.subnor.f64 
&UISETP.LT.U32.AND+2*UISETP.GE.U32.AND.EX
&8 \\
  \hline

\multicolumn{3}{|c|}{\textbf{copysign   Instruction}}
& \multicolumn{3}{|c|}{\textbf{Other   Instruction}}\\
  \hline
   copysign.f32  &  2*LOP3.LUT or 1.5*LOP3.LUT  & 4&sin.approx.f32 &FMUL + MUFU.SIN &8\\
  \hline
   copysign.f64 & 2*ULOP3.LUT+IMAD.U32+*MOV & 6&cos.approx.f32
 &FMUL.RZ+MUFU.COS  &8 \\
  \hline

\multicolumn{3}{|c|}{\textbf{and/or/xor    Instruction}}
&lg2.approx.f32  &FSETP.GEU.AND+FMUL+MUFU.LG2+FADD  &18\\
  \hline
   and.b16   &  LOP3.LUT  or 1.5*LOP3.LUT & 2& ex2.approx.f32 
 &FSETP.GEU.AND+2*FMUL+MUFU.EX2 &18\\
  \hline
   and.b32  &  LOP3.LUT & 2&ex2.approx.f16 &MUFU.EX2.F16 &6 \\
  \hline
   and.b64   &  ULOP3.LUT  & 2-3 &tanh.approx.f32   &MUFU.TANH &6 \\
  \hline

\multicolumn{3}{|c|}{\textbf{Not    Instruction}}
&tanh.approx.f16  &MUFU.TANH.F16  &6\\
  \hline
  not.b16  &  LOP3.LUT     & 2& bar.warp.sync; &NOP &changes\\
  \hline
   not.b32   &  LOP3.LUT & 2&fns.b32 &multiple  instructions &79 \\
  \hline
   not.b64&  2*ULOP3.LUT & 4 &cvt.rzi.s32.f32 &F2I.TRUNC.NTZ &6 \\
  \hline

\multicolumn{3}{|c|}{\textbf{lop3
    Instruction}}
&setp.ne.s32  &ISETP.NE.AND  &10\\
  \hline
  lop3.b32   & IMAD.MOV.U32+LOP3.LUT      & 4&mov.u32  clock&CS2R.32 &2  \\
  \hline
  
\multicolumn{3}{|c|}{\textbf{cnot     Instruction}} &\multicolumn{3}{|c|}{\textbf{Bfi     Instruction}}
\\
  \hline
  cnot.b16
  &  ULOP3.LUT+ISETP.EQ.U32.AND+SEL  & 5&bfi.b32  &  3*PRMT+2*IMAD.MOV+SHF.L.U32+BMSK+LOP3.LUT  & ~11\\
  \hline
   cnot.b32    &  UISETP.EQ.U32.AND+USEL  & 4 &bfi.b64
     &  UMOV+USHF.L.U32+(UIADD3+ULOP3.LUT)*  & 5\\
  \hline
  {cnot.b64} &  {multiple instructions} &  11  &\multicolumn{3}{|c|}{\textbf{dp4a.u32/s32   Instruction}}\\
  \hline

\multicolumn{3}{|c|}{\textbf{bfe      Instruction}}&dp4a.u32.u32   &  IMAD.MOV.U32+IDP.4A.U8.U8    & 135-170
\\
  \hline
  bfe.s32/.u32  &  3*PRMT+2*IMAD.MOV+SHF.R.U32.HI+SGXT/.U32  & 11 &\multicolumn{3}{|c|}{\textbf{dp2a.u32/s32   Instruction}}\\
  \hline
   bfe.u64     &  UMOV+USHF.L.U32+(UIADD3+ULOP3.LUT)*  & 5 &dp2a.lo.u32.u32    &  IMAD.MOV.U32+IDP.2A.LO.U16.U8&135-170\\
  \hline
   bfe.s64  &  multiple instructions  &  14 &&&\\
  \hline

  \end{tabular}}
   \label{table:ampere_cuda_11}
\end{table*}
%====================================

% \bibliographystyle{IEEEtran}
% \bibliography{references}

\clearpage
\begin{thebibliography}{50}
  
\bibitem{fb_supercomputer} 
  \emph{Meta AI Supercomputer,} 2022. [Online]. Available: \url{https://ai.facebook.com/blog/ai-rsc/}

\vspace{3pt}

\bibitem{top500}
  \emph{Top500 List.} [Online]. Available: \url{https://www.top500.org/lists/top500/2022/06/}

\vspace{3pt}

\bibitem{zhang2017understanding}
X. Zhang, G. Tan, S. Xue, J. Li, K. Zhou, and M. Chen, “Understanding
the gpu microarchitecture to achieve bare-metal performance tuning," in \emph{Proceedings of the 22nd ACM SIGPLAN Symposium on Principles and
Practice of Parallel Programming,} 2017, pp. 31–43.

\vspace{3pt}

\bibitem{ho2017exploiting}
N.-M. Ho and W.-F. Wong, “Exploiting half precision arithmetic in
nvidia gpus," in \emph{2017 IEEE High Performance Extreme Computing
Conference (HPEC),} IEEE, 2017, pp. 1–7.

\vspace{3pt}

\bibitem{lustig2019formal}
D. Lustig, S. Sahasrabuddhe, and O. Giroux, “A formal analysis of
the nvidia ptx memory consistency model," in \emph{Proceedings of the
Twenty-Fourth International Conference on Architectural Support for
Programming Languages and Operating Systems,} 2019, pp. 257–270.

\vspace{3pt}

\bibitem{wong2010demystifying}
H. Wong, M.-M. Papadopoulou, M. Sadooghi-Alvandi, and  A. Moshovos, “Demystifying gpu microarchitecture through microbenchmarking," in \emph{2010 IEEE International Symposium on Performance Analysis of Systems and Software (ISPASS),} IEEE, 2010, pp. 235–246.

\vspace{3pt}

\bibitem{jia2018dissecting}
Z. Jia, M. Maggioni, B. Staiger, and D. P. Scarpazza, “Dissecting the
nvidia volta gpu architecture via microbenchmarking," \emph{arXiv preprint,} arXiv:1804.06826, 2018.


\vspace{3pt}

\bibitem{jia2019dissecting}
Z. Jia, M. Maggioni, J. Smith, and D. P. Scarpazza, “Dissecting
the nvidia turing t4 gpu via microbenchmarking," \emph{arXiv preprint,} arXiv:1804.06826, 2019.


\vspace{3pt}

\bibitem{van2022isolating}
R. van Stigt, S. N. Swatman, and A.-L. Varbanescu, “Isolating gpu
architectural features using parallelism-aware microbenchmarks," in \emph{Proceedings of the 2022 ACM/SPEC on International Conference on
Performance Engineering,} 2022, pp. 77–88.

\vspace{3pt}

\bibitem{andersch2015latency}
M. Andersch, J. Lucas, M. A. LvLvarez-Mesa, and B. Juurlink, “On latency in gpu throughput microarchitectures," in \emph{2015 IEEE International
Symposium on Performance Analysis of Systems and Software (ISPASS),} IEEE, 2015, pp. 169–170.

\vspace{3pt}

\bibitem{Nvidia_ampere_arch_doc}
\emph{NVIDIA A100 Tensor Core GPU Architecture,} 2022. [Online]. Available: \url{https://images.nvidia.com/aem-dam/en-zz/Solutions/datacenter/nvidia-ampere-architecture-whitepaper.pdf}.

\vspace{3pt}

\bibitem{Nvidia_PTX}
\emph{Parallel Thread Execution ISA Version 7.7,} 2022. [Online]. Available: \url{https://docs.nvidia.com/cuda/parallel-threadexecution/index.htmlinstruction-set}.

\vspace{3pt}

\bibitem{arafa2019low}
Y. Arafa, A.-H. A. Badawy, G. Chennupati, N. Santhi, and S. Eidenbenz, “Low overhead instruction latency characterization for nvidia gpgpus," in \emph{2019 IEEE High Performance Extreme Computing Conference (HPEC),} IEEE, 2019, pp. 1–8.

\vspace{3pt}

\bibitem{volkov2018microbenchmark}
V. Volkov, “A microbenchmark to study gpu performance models," in \emph{ACM
SIGPLAN Notices,} vol. 53, no. 1, pp. 421–422, 2018.

\vspace{3pt}

\bibitem{bakhoda2009analyzing}
A. Bakhoda, G. L. Yuan, W. W. Fung, H. Wong, and T. M. Aamodt, “Analyzing cuda workloads using a detailed gpu simulator," in \emph{2009
IEEE international symposium on performance analysis of systems and
softwar,}  IEEE, 2009, pp. 163–174.

\vspace{3pt}

\bibitem{samadi2014paraprox}
M. Samadi, D. A. Jamshidi, J. Lee, and S. Mahlke, “Paraprox: Patternbased approximation for data parallel applications," in \emph{Proceedings of the
19th international conference on Architectural support for programming
languages and operating systems,} 2014, pp. 35–50.

\vspace{3pt}

\bibitem{arafa2021hybrid}
Y. Arafa, A.-H. Badawy, A. ElWazir, A. Barai, A. Eker, G. Chennupati,
N. Santhi, and S. Eidenbenz, “Hybrid, scalable, trace-driven performance
modeling of gpgpus," in \emph{Proceedings of the International Conference for
High Performance Computing, Networking, Storage and Analysis,}  2021,
pp. 1–15.


\vspace{3pt}

\bibitem{arafa2019ppt}
Y. Arafa, A.-H. A. Badawy, G. Chennupati, N. Santhi, and S. Eidenbenz, “Ppt-gpu: Scalable gpu performance modeling," \emph{IEEE Computer
Architecture Letters,} vol. 18, no. 1, pp. 55–58, 2019.

\vspace{3pt}

\bibitem{arafa2020verified}
Y. Arafa, A. ElWazir, A. ElKanishy, Y. Aly, A. Elsayed, A.-H. Badawy,
G. Chennupati, S. Eidenbenz, and N. Santhi, “Verified instruction-level
energy consumption measurement for nvidia gpus," in \emph{Proceedings of
the 17th ACM International Conference on Computing Frontiers,} 2020,
pp. 60–70.

\vspace{3pt}

\bibitem{mei2016dissecting}
X. Mei and X. Chu, “Dissecting gpu memory hierarchy through
microbenchmarking," \emph{IEEE Transactions on Parallel and Distributed
Systems,} vol. 28, no. 1, pp. 72–86, 2016.

\vspace{3pt}

\bibitem{sun2022dissecting}
W. Sun, A. Li, T. Geng, S. Stuijk, and H. Corporaal, “Dissecting
tensor cores via microbenchmarks: Latency, throughput and numerical
behaviors," \emph{arXiv preprint arXiv:2206.02874,} 2022.

\vspace{3pt}

\bibitem{markidis2018nvidia}
S. Markidis, S. W. Der Chien, E. Laure, I. B. Peng, and J. S. Vetter, “Nvidia tensor core programmability, performance \& precision," in \emph{2018 IEEE international parallel and distributed processing symposium
workshops (IPDPSW),} IEEE, 2018, pp. 522–531.

\vspace{3pt}

\bibitem{fasi2021numerical}
M. Fasi, N. J. Higham, M. Mikaitis, and S. Pranesh, “Numerical behavior
of nvidia tensor cores," \emph{PeerJ Computer Science,} vol. 7, p. e330, 2021.

\vspace{3pt}

\bibitem{bombieri2016mipp}
N. Bombieri, F. Busato, F. Fummi, and M. Scala, “Mipp: A microbenchmark suite for performance, power, and energy consumption
characterization of gpu architectures," in \emph{2016 11th IEEE Symposium
on Industrial Embedded Systems (SIES),} IEEE, 2016, pp. 1–6.

\vspace{3pt}

\bibitem{raihan2019modeling}
M. A. Raihan, N. Goli, and T. M. Aamodt, “Modeling deep learning
accelerator enabled gpus," in \emph{2019 IEEE International Symposium on
Performance Analysis of Systems and Software (ISPASS),} IEEE, 2019,
pp. 79–92.

\vspace{3pt}

\bibitem{yan2020demystifying}
D. Yan, W. Wang, and X. Chu, “Demystifying tensor cores to optimize
half-precision matrix multiply," in \emph{2020 IEEE International Parallel and
Distributed Processing Symposium (IPDPS),} IEEE, 2020, pp. 634–643.

\vspace{3pt}

\bibitem{kothiya2014understanding}
M. V. Kothiya et al, “Understanding the isa impact on gpu architecture," 2014.

\vspace{3pt}

\bibitem{Nvidia_Cuda}
\emph{NVIDIA CUDA programming. User’s guide,} 2022. [Online]. Available: \url{https://docs.nvidia.com/cuda/cuda-c-programming-guide/index.html}.

\end{thebibliography}

\end{document}